\documentclass[aps,pre,twocolumn,groupedaddress]{revtex4-1}
\usepackage{graphicx}
\usepackage{dcolumn}
\usepackage{bm}
\usepackage{amsmath}
\usepackage{amssymb}
\usepackage{color}

\begin{document}

\title{Excitable reaction-diffusion waves of curvature-inducing proteins on deformable membrane tubes}

\author{Naoki Tamemoto}
\affiliation{Institute for Solid State Physics, University of Tokyo, Kashiwa, Chiba 277-8581, Japan}
\author{Hiroshi Noguchi}
\email[]{noguchi@issp.u-tokyo.ac.jp}
\affiliation{Institute for Solid State Physics, University of Tokyo, Kashiwa, Chiba 277-8581, Japan}


\begin{abstract}
Living cells employ excitable reaction-diffusion waves for internal cellular functions, in which curvature-inducing proteins are often involved. However, the role of their mechanochemical coupling
is not well understood.
Here, we report the membrane deformation induced by the excitable reaction-diffusion waves of curvature-inducing proteins
and the alternation in the waves due to the deformation, using a coarse-grained simulation of tubular membranes with a modified  FitzHugh--Nagumo model.
Protein-propagating waves deform tubular membranes, and large deformations induce budding and erase waves.
The wave speed and shape are determined by a combination of membrane deformation and spatial distribution of the curvature-inducing protein.
Waves are also undulated in the azimuthal direction depending on the condition.
Rotationally symmetric waves locally deform the tubes into a symmetric shape but maintain a straight shape on average.
Our simulation method can be applied to other chemical reaction models and used to investigate various biomembrane phenomena.
\end{abstract}

\maketitle

\section{Introduction}
Membrane deformation is a common phenomenon in living cells \cite{Zimmerberg1,McMahon2,Suetsugu1,Voeltz1}.
The shapes of biomembranes are regulated by various proteins including curvature-inducing proteins \cite{Zimmerberg1,McMahon2,Suetsugu1,Baumgart2}. 
For example, Bin/Amphiphysin/Rvs (BAR) superfamily proteins, which have a banana-shaped dimer structure, bind to membranes and bend them according to intrinsic protein shapes \cite{Suetsugu1,Baumgart2,PeterBJ1}. Clathrin, and coat protein complexes (COPI and COPII) also bend membranes and induce membrane budding.
These proteins cooperatively work.
For example, during clathrin-mediated endocytosis,
the BAR proteins bind to membranes, and subsequently, clathrin-assembly forms a spherical bud; the binding of dynamin to the bud neck induces membrane fission \cite{McMahon1}.
While proteins bend membranes as described above, the membrane curvature affects protein binding (curvature sensing)~\cite{McMahon2,Baumgart2,PeterBJ1,ZhaoW1,Sorre1,Has1}. 
Various curvature-inducing proteins have different preferences for high membrane curvatures depending on protein shape. Thus, membrane curvature also affects their dynamics \cite{ZhaoW1}. 

Recently, there have been experimental and theoretical studies on the feedback regulation between protein dynamics and membrane mechanics.
Reaction-diffusion waves of proteins coupled with membrane deformation have been observed and theoretically analyzed under various conditions \cite{WuZ1,WuM1,Peleg1,GovNS2}.
Further, the membrane shape affects intracellular proteins' wave propagation, and they subsequently regulate cell motion \cite{Weiner1,Taniguchi2,Miao1}.
The dynamics of the curvature-inducing proteins are associated with cellular functions, such as cell migration \cite{Tsujita1}, and cell division \cite{XiaoS1}.
In addition, giant unilamellar vesicles (GUVs) containing Min proteins, which generate reaction-diffusion waves in {\it Escherichia coli} and regulate cell division, exhibit cyclic deformation accompanying the Min protein waves \cite{Godino1,Christ1,Litschel1}.
In signal transduction, receptors on cell membranes receive extracellular signals, transfer information downstream in the intracellular reaction network, and regulate cellular functions according to the surrounding environment, such as chemotaxis \cite{Xiong1}. Some of these cell signaling are modeled using excitable reaction-diffusion models, and various studies have shown that the signaling network regulates cell migration \cite{Huang1,ZhanW1}.
Researchers have analyzed excitable reaction-diffusion waves on fixed curved surfaces and found that the surface geometry affects the stability, speed, and shape of the waves \cite{Horibe1,Kneer1}. 
However, studies that have considered membrane deformation and reaction-diffusion dynamics are limited, and their biological implications are not understood yet.

In previous studies, we examined the coupling between the large deformation of a fluid vesicle and reaction-diffusion dynamics with the binding of curvature-inducing proteins using a dynamically triangulated membrane model and a non-excitable reaction-diffusion model, Brusselator model, to investigate the mechanochemical feedback \cite{Tamemoto1,Tamemoto2}. 
We understood that membrane deformation stabilizes Turing patterns, simultaneously Turing patterns induce membrane deformation, resulting in the formation of budded and multispindle-shaped vesicles \cite{Tamemoto1}.
Moreover, we successfully reproduced the self-oscillation of vesicle shapes as seen in the reconstructed Min system in GUVs, 
and showed that the starting point of traveling waves is determined by the local membrane curvature~\cite{Tamemoto2}.

In this study, we simulated a coupled system between membrane deformation and excitable reaction-diffusion waves of curvature-inducing proteins. 
Excitable reaction-diffusion waves are fundamental biological phenomena, and models, such as the FitzHugh--Nagumo model for nerve signaling, have been used for studying them \cite{FitzHugh1,Nagumo1}.
Moreover, various protein waves interacting with membrane mechanics are demonstrated using excitable reaction-diffusion models \cite{WuZ1,WuM1}.
We employed the modified FitzHugh--Nagumo model to be coupled with membrane deformation.
Here, we consider that the curvature-inducing proteins have a laterally isotropic shape.
We employed tubular membranes and waves propagating along the tube axis to reduce the complexity of waves with higher dimensions.
Tubular membrane structures are observed in cell membranes (e.g., axons of neurons) as well as intracellular organelles (e.g., endoplasmic reticulum (ER) and mitochondria)~\cite{McMahon2,Voeltz1}.
Tubular membranes can be produced in vitro by pulling a vesicle using optical tweezers and a micropipette,
and the tube radius is controlled by the amplitude of the imposed force~\cite{Dimova1,Roux1}. Experimentally, tubular membranes are often used to investigate the curvature dependence of protein binding to membranes \cite{Baumgart2,Sorre1,Has1}.
We investigated the dynamics of protein-propagating waves using a combination of the FitzHugh--Nagumo model and tubular membranes. 

\section{Methods}
\subsection{Membrane model}
We applied the membrane model used in our previous studies \cite{Tamemoto1,Tamemoto2} with a few modifications.
We triangulated a membrane surface to discretize the membrane tube. In this triangular network, $N$ vertices are connected by bonds of average length $a$, and the vertices have excluded volumes and masses $m$. The surface area $S$ is constrained by the harmonic potentials \cite{Noguchi3}.
The tube lies along the $x$-axis and is connected using a periodic boundary condition. Unlike our previous studies, the volume enclosed by the membrane is left unconstrained in this study. This tube is considered to be extended from a vesicle by optical tweezers or protein filaments, so that
the tube is connected to a reservoir of a large volume.
The triangulated bond network is re-meshed using the bond-flip Monte Carlo method to allow lateral diffusion of vertices \cite{Gompper1}. 

The deformation of the tubular membrane is expressed by free energy of the curvature, which is modified to include the effect of curvature-inducing proteins. 
The curvature energy is expressed as $F_{\mathrm{cv}} = \int f_{\mathrm{cv}} dS$ with
\begin{equation}
f_{\mathrm{cv}} = (1-u) \frac{\kappa_0}{2} (2H)^2 + u \frac{\kappa_1}{2} (2H-C_0)^2, \label{Eq:modCurvE}
\end{equation}
where $u$ is the concentration of curvature-inducing proteins on the membranes ($u \in [0,1]$); $H$ is the mean curvature; $\kappa_1$ and $\kappa_0$ are the bending rigidity with and without the proteins on the membranes, respectively; and $C_0$ is the spontaneous curvature of the curvature-inducing proteins \cite{Tamemoto1,Tamemoto2,Noguchi6,Noguchi9}. 
Although the difference in the saddle-splay modulus $\bar{\kappa}$ (also called the Gaussian modulus) can modify the protein binding~\cite{Noguchi6,Noguchi9}, it is not considered here for simplicity.
For the budding of a vesicle, the $\bar{\kappa}$ difference can be accounted for by the rescaling of $\kappa_1$ and $C_0$~\cite{Noguchi6}. Similarly, the $\bar{\kappa}$ difference does not likely cause  a qualitative change in the present case.
When $\bar{\kappa}$ is constant for an entire membrane,
$\bar{\kappa}$ does not contribute to membrane shape transformation for a fixed topology because of the Gauss--Bonnet theorem.
According to this curvature energy expression, the larger the protein concentration $u$, the more strongly the membrane bends when $C_0$ is sufficiently large.

Membrane deformation is solved through molecular dynamics simulation using a Langevin thermostat:
\begin{equation}
  m \frac{\partial^2 \bm{r}_i}{\partial t^2} = - \frac{\partial U}{\partial \bm{r}_i} - \zeta \frac{\partial \bm{r}_i}{\partial t} + \bm{g}_i(t), \label{Eq:MD}
\end{equation}
where $\zeta$ is the friction coefficient and $\bm{g}_i$ is Gaussian white noise, which obeys the fluctuation-dissipation theorem. We use the potential $U = U_{\mathrm{S}} + U_{\mathrm{b}} + U_{\mathrm{r}} + U_{\mathrm{cv}}$, where $U_{\mathrm{S}}$ is the constraint potential for the surface area $S$; $U_{\mathrm{b}}$ and $U_{\mathrm{r}}$ are the bond and repulsive potentials, respectively; and $U_{\mathrm{cv}}$ is the discretized potential for the bending energy, $F_{\mathrm{cv}}$, using the dual lattice. 
\begin{eqnarray}
U_{\mathrm{S}} &=& \frac{1}{2}k_{\mathrm{S}}(S-S_0)^2,\\
U_{\mathrm{b}} &=& \sum_{\mathrm{bond}} \frac{b\exp(1/(l_{\mathrm{c0}}-r_{i,j}))}{l_{\mathrm{max}}-r_{i,j}}\Theta(r_{i,j} - l_{\mathrm{c0}}), \\
U_{\mathrm{r}} &=& \sum_{\mathrm{all\ pairs}} \frac{b \exp(1/(r_{i,j}-l_{\mathrm{c1}}))}{r_{i,j}-l_{\mathrm{min}}}\Theta(l_{\mathrm{c1}} - r_{i,j}), \\
U_{\mathrm{cv}} &=& (1-u)\frac{\kappa_0}{2} \sum_i \frac{1}{\sigma_i}
  \bigg(  \sum_{j(i)} \frac{\sigma_{i,j}{\bf r}_{i,j}}{r_{i,j}}\bigg)^2 \\
&&+ u\frac{\kappa_1}{2} \sum_i \sigma_i  
  \bigg( \frac{1}{\sigma_i} \sum_{j(i)} \frac{\sigma_{i,j}{\bf r}_{i,j}}{r_{i,j}} - C_0{\bf n}_i \bigg)^2,\nonumber
\end{eqnarray} 
where $\Theta(x)$ is the unit step function, $r_{i,j}$ is the distance between two vertices $i$ and $j$,
and ${\bf n}_i$ is the normal vector at the $i$-th vertex.
The sum over $j(i)$ is over the neighbors of the  $i$-th vertex,  which are connected by tethers.
The length of a bond in the dual lattice is $\sigma_{i,j}=r_{i,j}[\cot(\theta_1)+\cot(\theta_2)]/2$, 
where the angles $\theta_1$ and $\theta_2$ are opposite to bond $ij$ in 
the two triangles sharing this bond. 
$\sigma_i=0.25\sum_{j(i)} \sigma_{i,j}r_{i,j}$ is the area of the dual cell of vertex $i$.
Here, we use $k_{\mathrm{S}}= 4k_{\mathrm{B}}T$, $b=80k_{\mathrm{B}}T$, $l_{\mathrm{max}}=1.33a$, $l_{\mathrm{c0}}=1.15a$, $l_{\mathrm{c1}}=0.85a$, and $l_{\mathrm{min}}=0.67a$.
The details of these potentials are provided in Ref.~\cite{Noguchi3}. 

\subsection{Reaction-diffusion model}
\begin{figure}[!tb]
  \centering
  \includegraphics[width=8cm]{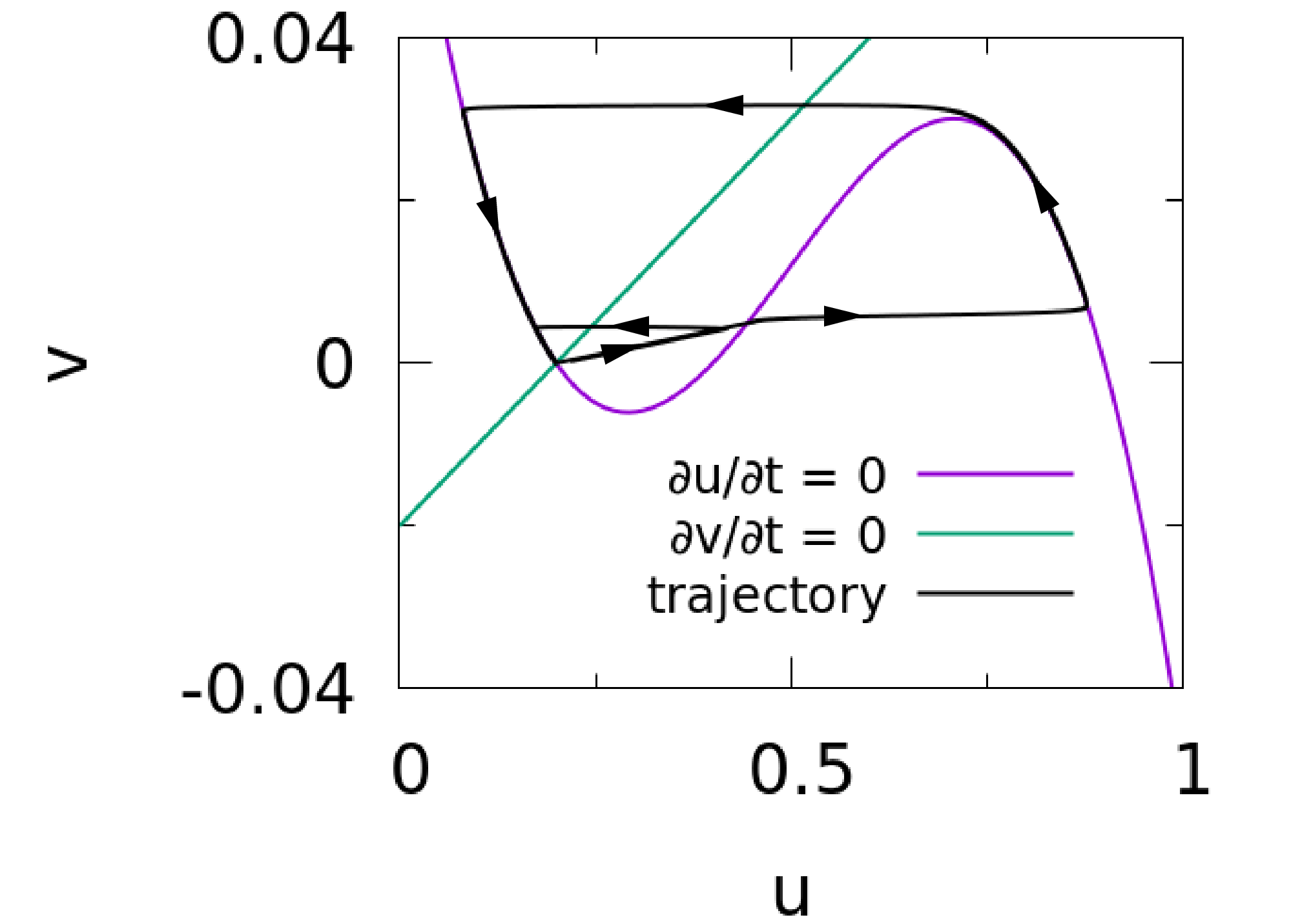}
  \caption{\label{Fig:nullcline}
    Phase plane of the FitzHugh--Nagumo model with the modified protein unbinding process for $A=0.4$, $B = 0.001$, $C=0.1$, and $G=0$. Purple and green lines are nullclines for $\partial u/\partial t=0$ and $\partial v/\partial t=0$, respectively. Black lines indicate trajectories after the equilibrium state is stimulated.
  }
\end{figure}

We employed the FitzHugh--Nagumo model~\cite{FitzHugh1,Nagumo1} as an excitable reaction-diffusion model to investigate stable propagating waves with excitable dynamics.
Reaction-diffusion equations for concentrations of curvature-inducing proteins $u$ and regulatory proteins $v$ are given by
\begin{align}
  \begin{split}
    \tau \frac{\partial u}{\partial t} & = -(u-0.2)(u-A)(u-0.9) - v - G \frac{\partial f_{\mathrm{cv}}}{\partial u} \\
    & \quad + D_u \nabla^2 u + Q_u, \label{Eq:FHNCV.u}
  \end{split} \\
  \tau \frac{\partial v}{\partial t} & = B(C(u-0.2)-v) + D_v \nabla^2 v + Q_v, \label{Eq:FHNCV.v}
\end{align}
where $\tau$ is a time constant; $D_u$ and $D_v$ are the diffusion coefficients of the two types of proteins; $A$, $B$, and $C$ are positive reaction parameters; $Q_u$ and $Q_v$ are the stimuli for wave initiation at a thin slice of the tube; and $G$ is the magnitude of mechanochemical coupling.
To introduce the effect of membrane curvature on the reaction-diffusion model, we assumed that the binding rate of curvature-inducing proteins onto membranes is proportional to the partial derivative of the local curvature energy with respect to the concentration of curvature-inducing proteins, $\partial f_{\mathrm{cv}}/\partial u$, as in our previous study \cite{Tamemoto1}. According to this coupling, the binding of curvature-inducing proteins is enhanced around the preferred curvature $H \simeq C_0/2$, where $\partial f_{\mathrm{cv}}/\partial u < 0$.
We constrain the range of $u$ during the time evolution to maintain $u \in [0,1]$;
when Eq.~(\ref{Eq:FHNCV.u}) gives $u$ out of this range, it is set to $u=0$ or $u=1$.
When a small stimulus is given in the equilibrium state, the state quickly recovers to the fixed point. However, the system shows excitability when a larger stimulus is given (see Fig.~\ref{Fig:nullcline}).

\subsection{Parameters}
In this study, we used membrane tubes with a length of $L_x=200a$, where the number of vertices is $N=15~000$ or $30~000$. 
For these numbers of vertices, the tubular surface areas are $S\simeq 12~300a^2$ and $24~600a^2$, and the corresponding radii of the tubes are $R_{\mathrm{t}}=S/2\pi L_x \simeq 9.79a$ and $19.6a$, respectively. Unless otherwise mentioned, we applied the conditions $\kappa_0/k_\mathrm{B} T=20$, $A=0.4$, $B=0.001$, $C=0.1$, $D_u/a^2=0.1$, $D_v=0$, $G k_\mathrm{B} T/a^2=0.002$, and $\tau_\mathrm{md}=\tau$, where $\tau_\mathrm{md}=\zeta a^2/k_\mathrm{B}T$. 
For this ratio of the time units of the reaction and membrane motion, the membrane motion can follow the reaction dynamics. 
The starting protein concentration values were set around the fixed point calculated using a fixed value of the membrane curvature $(H=1/2R_{\mathrm{t}})$, with small, random perturbations.
			
\begin{figure}[!tb]
  \centering
  \includegraphics[width=8cm]{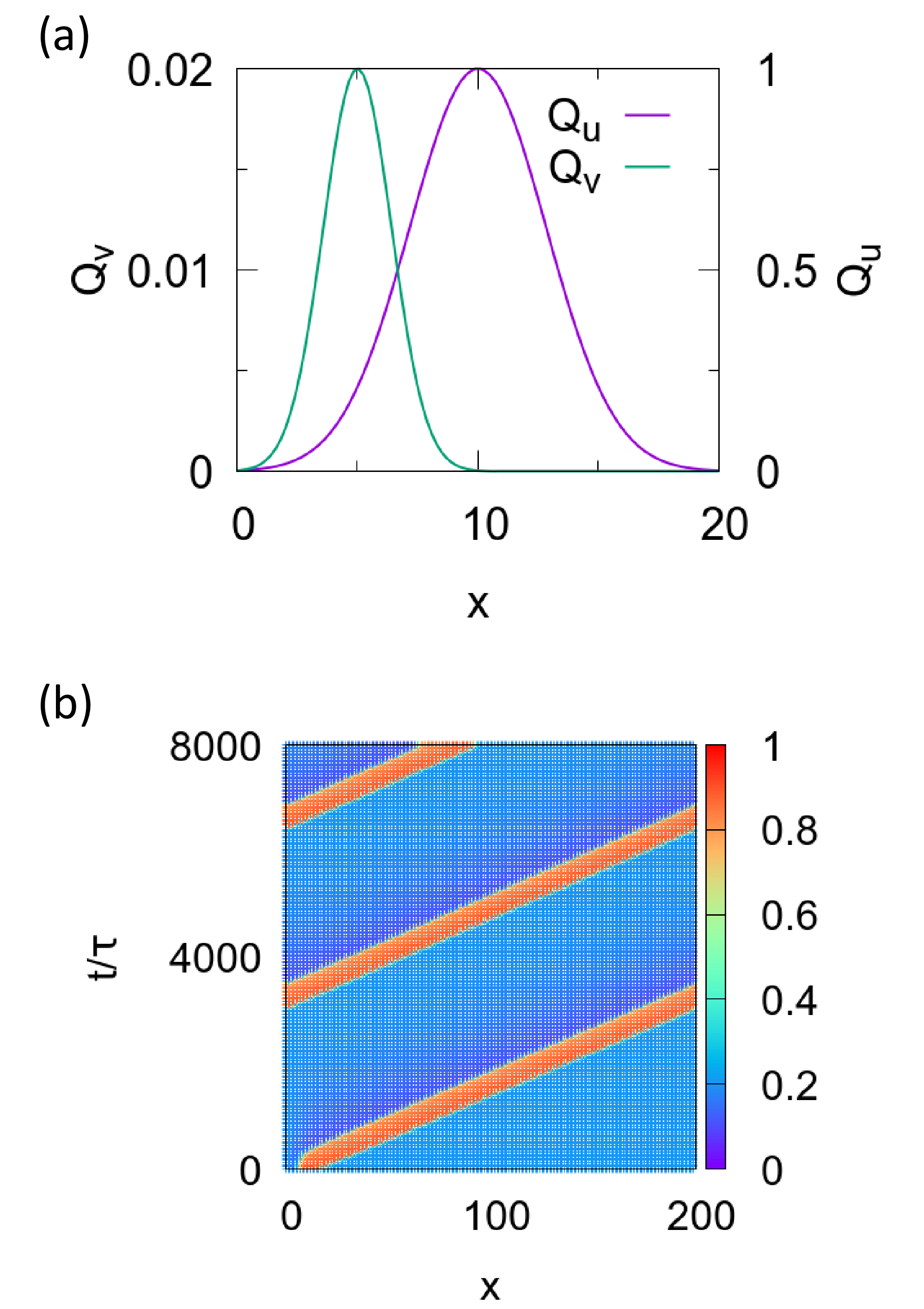}
  \caption{\label{Fig:1Dwave}
    Initiation of propagating waves. (a) The profiles of the stimuli $Q_u$ and $Q_v$ to initiate the propagating wave. 
(b) Propagating waves induced by the stimuli shown in (a). The color variation indicates different concentration of the curvature-inducing protein, $u$.
  }
\end{figure}

For the initiation of propagating waves, we added stimuli at the thin slice of the tubular membrane,
given by
\begin{align}
  Q_u & = \exp \left(-0.0625(x-10)^2 \right) \text{ and} \label{Eq:st.u} \\
  Q_v & = 0.02\exp \left(-0.25(x-5)^2 \right) \label{Eq:st.v}
\end{align}
during the initial short period ($0 \leq t/\tau \leq 1$).
The profiles of $Q_u$ and $Q_v$ are shown in Fig.~\ref{Fig:1Dwave}(a), 
and they induce a one-dimensional wave that propagates on a flat membrane with $G=0$ as shown in Fig.~\ref{Fig:1Dwave}(b).

Equations~(\ref{Eq:FHNCV.u}) and (\ref{Eq:FHNCV.v}) were integrated numerically using the forward difference method for time and the finite-volume scheme for space \cite{Tamemoto1}.
Equation~(\ref{Eq:MD}) was integrated using the leapfrog method.
Error bars were calculated based on three independent runs.

In the estimation of local curvature, we calculated a smoothed local curvature $\tilde{H}$ by averaging the local curvature $H$ up to the second-order adjacent vertices. The details are described in the Supplementary Material of Ref.~\cite{Tamemoto1}.

\subsection{Stability of Nonexcited Membrane}
In thermal equilibrium,
a cylindrical membrane tube generates an force $f_{\mathrm{ex}}$ along the tube axis~\cite{Noguchi7}:
\begin{equation} \label{eq:force}
f_{\mathrm{ex}} 
 =  2\pi \kappa_0C_0\bigg\{\frac{1}{R_{\mathrm{t}}C_0}\Big[\Big(\frac{\kappa_1}{\kappa_0}-1\Big)u + 1\Big]  -  \frac{\kappa_1}{\kappa_0}u \bigg\}.
\end{equation}
For $f_{\mathrm{ex}} > 0$, the membrane tube is stable and this force is balanced with an external force (here imposed by the periodic boundary condition).
In contrast, for $f_{\mathrm{ex}} \leq 0$, the cylindrical tubes are unstable, so that
the unstable condition is given by
\begin{equation} \label{eq:fzero}
C_0R_{\mathrm{t}}  \geq  1 + \frac{\kappa_0(1-u)}{\kappa_1 u} .
\end{equation}
This condition is identical to that in Ref.~\cite{Zhong1} (the membrane spontaneous curvature $\geq$ the tube curvature).
In particular, $f_{\mathrm{ex}}=0$, 
the membrane deforms spherical buds via unduloid deformation, maintaining the mean curvature of the membrane everywhere~\cite{Kenmotsu1,Naito1}.
Previously, budding of a tubular membrane with a positive spontaneous curvature has been reported~\cite{Tsafrir1,Campelo1,Tozzi1,Noguchi7}.
Note that the budding also occurs slightly below this threshold under thermal fluctuations ~\cite{Noguchi7}.
Buckling of the tube is induced by a large excess membrane area when the tube is compressed (negative surface tension)~\cite{LiuX1,denOtter1}.
In our simulation condition, an unexcited membrane has $u\simeq 0.2$ (see Fig.~\ref{Fig:nullcline}).
For example, the cylindrical tube is unstable at $C_0R_{\mathrm{t}} \gtrsim 3$ for $\kappa_1/\kappa_0=2$.

\section{Results and Discussions}
\subsection{Propagating waves with mechanochemical coupling}
\begin{figure}[!tb]
  \centering
  \includegraphics[width=8cm]{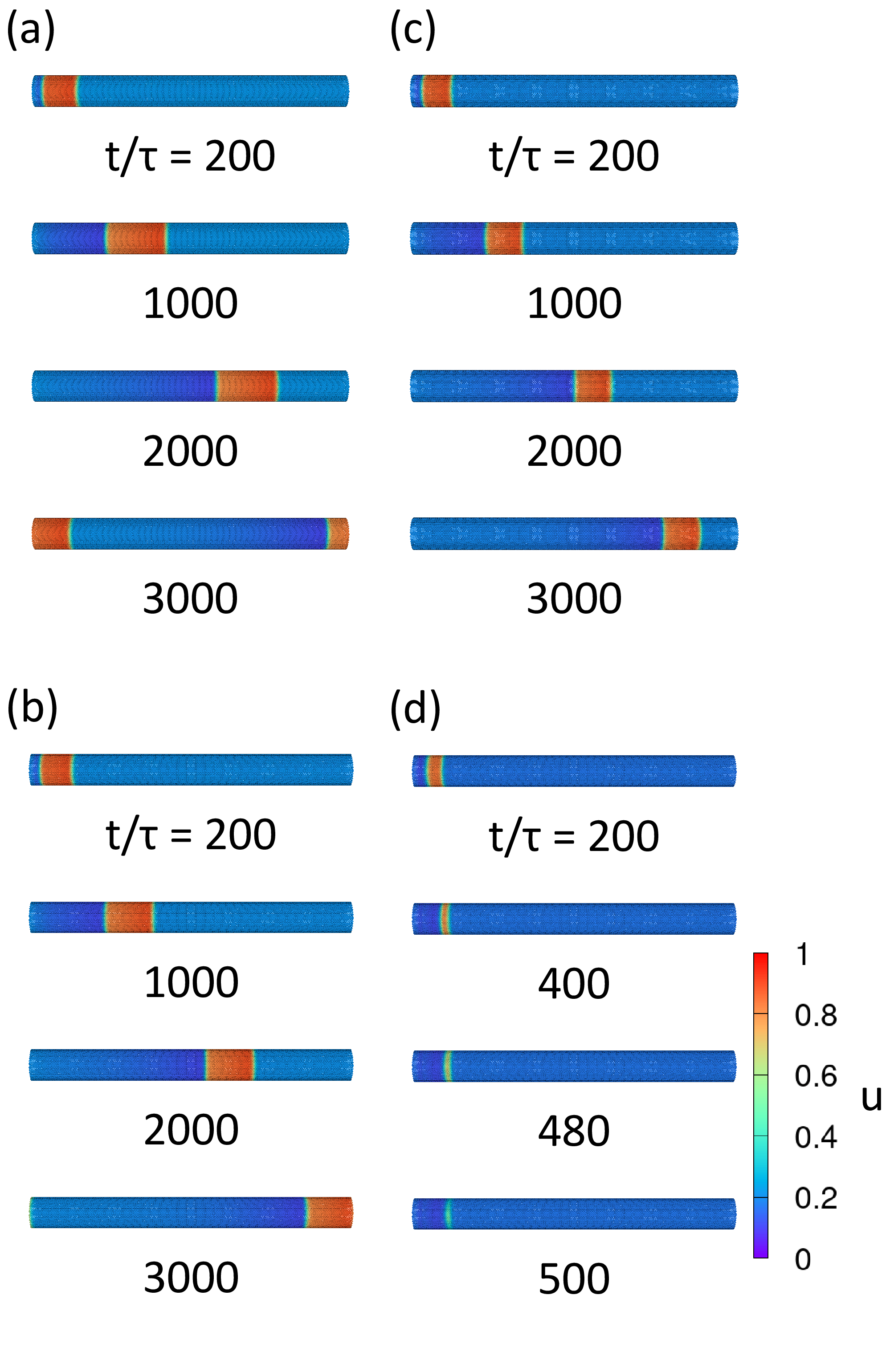}
  \caption{\label{Fig:tube_snap0}
    Typical sequential snapshots of propagating waves on non-deformable tubular membranes for (a) $I = 0.002$, (b) $I = 0$, (c) $I = -0.002$, and (d) $I=-0.01$. The color variation indicates different concentration of the curvature-inducing protein, $u$.
  }
\end{figure}

\begin{figure}[!tb]
  \centering
  \includegraphics[width=8cm]{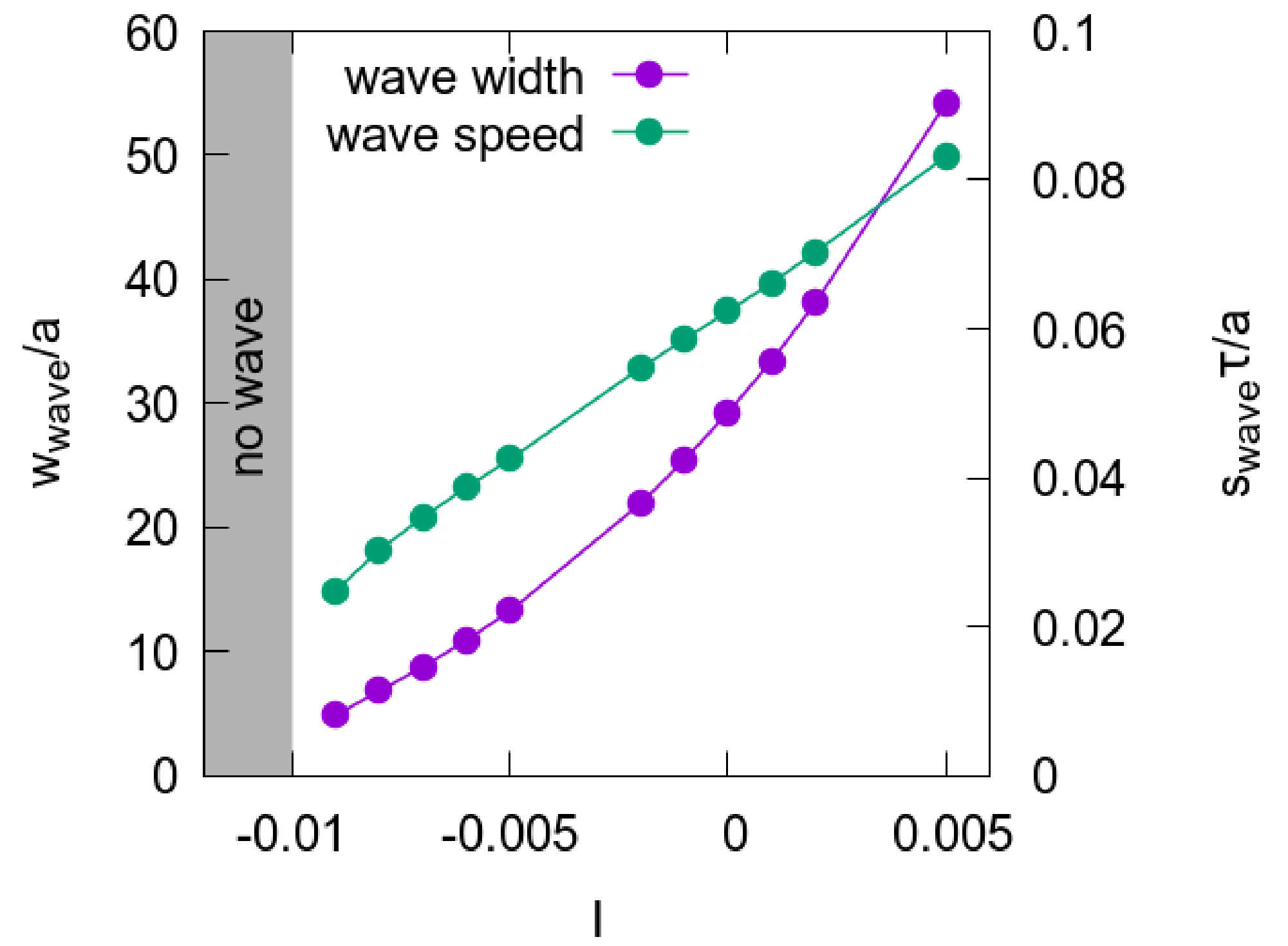}
  \caption{\label{Fig:3Dwave}
    Effect of constant inputs $I$ on the wave properties with non-deformable membranes. In the gray shaded area, propagating waves disappear.
  }
\end{figure}

We investigated the propagating waves on tubular membranes for $R_{\mathrm{t}}= 9.79a$ and $19.6a$ (narrow tubes and wide tubes, respectively). 
Before examining the deformable tubes, we evaluated the effects of constant positive or negative inputs on waves on a non-deformable narrow membrane tube. 
Instead of $-G\partial f_{\mathrm{cv}}/\partial u$, we added a constant input $I$ to Eq.~(\ref{Eq:FHNCV.u}) over the entire membrane-tube region. Figures~\ref{Fig:tube_snap0}(a)--(c) show typical snapshots of the propagating waves for $I=0.002$, $0$, and $-0.002$, respectively. Among them, the width and speed of the waves were largest for $I=0.002$ and smallest for $I=-0.002$. 
The wave is generated once but shrinks and disappears without propagating for $I=-0.01$ (see Fig.~\ref{Fig:tube_snap0}(d)). Figure~\ref{Fig:3Dwave} shows the changes in wave width, $w_{\mathrm{wave}}$ and speed, $s_{\mathrm{wave}}$. Both of wave speed and width decrease with decreasing $I$, 
and the waves disappear when $I$ is smaller than the threshold.

\begin{figure}[!tb]
  \centering
  \includegraphics[width=8cm]{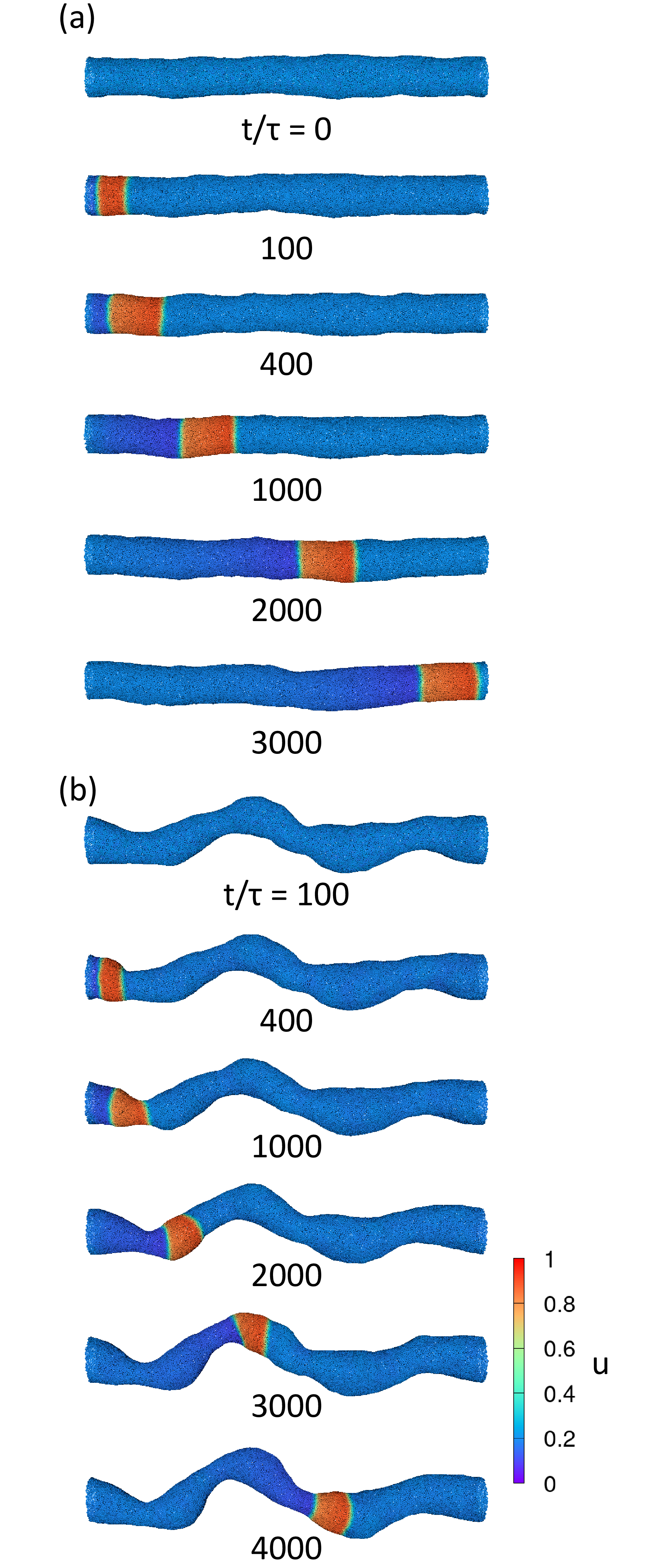}
  \caption{\label{Fig:tube_snap1}
    Typical sequential snapshots of propagating waves on deformable tubular membranes with a small radius ($R_{\mathrm{t}}=9.79a$)  and small spontaneous curvatures $C_0$.
(a) $C_0R_{\mathrm{t}}=1$ and $\kappa_1/\kappa_0=2$. (b) $C_0R_{\mathrm{t}}=3$ and $\kappa_1/\kappa_0=3$. The color variation indicates different concentration of the curvature-inducing protein, $u$.
The corresponding movies are provided in Supplementary Movies S1 and S2.  }
\end{figure}

\begin{figure}[!tb]
  \centering
  \includegraphics[width=8cm]{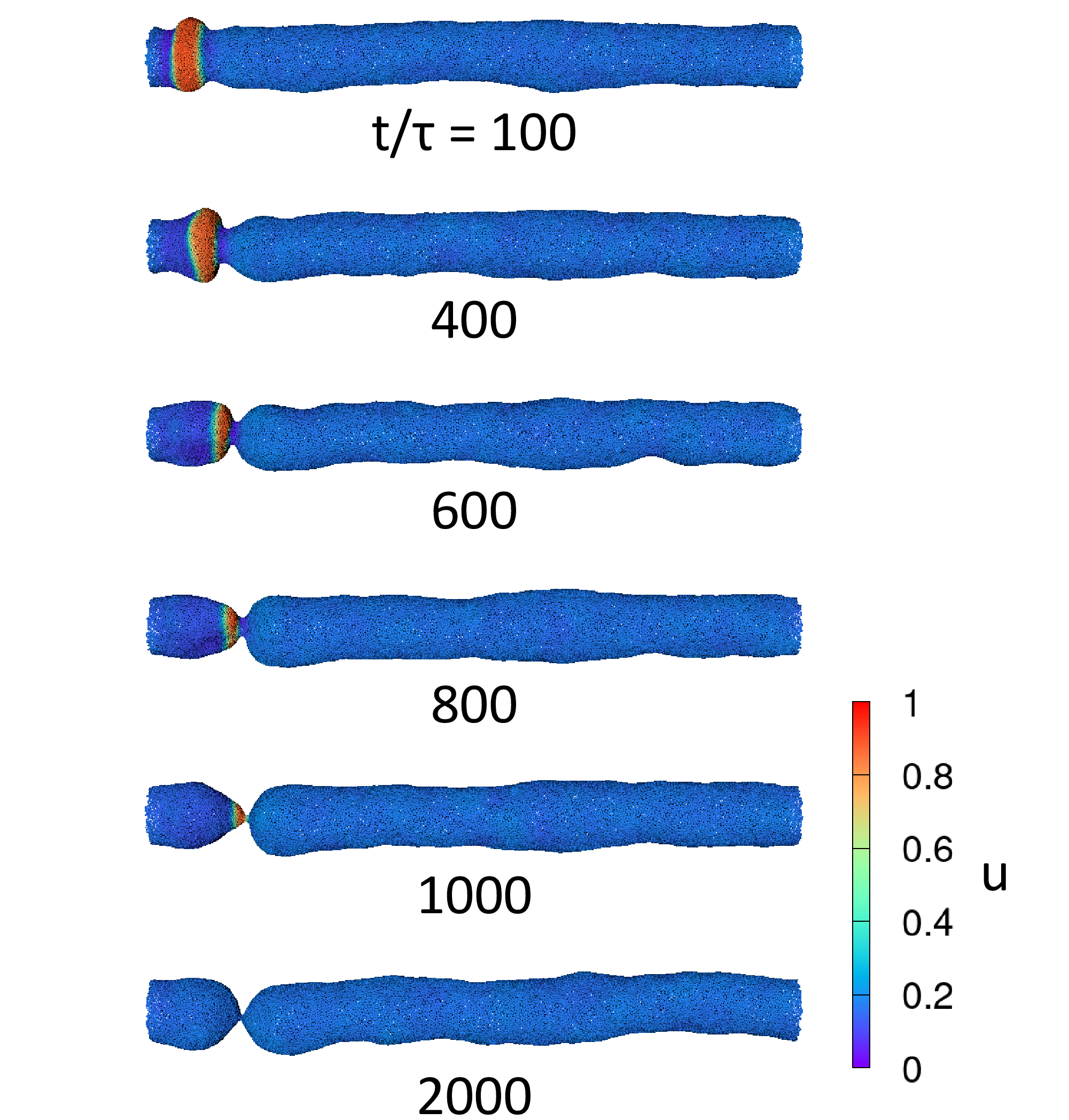}
  \caption{\label{Fig:tube_snap2}
    Typical sequential snapshots of propagating waves on deformable tubular membranes with the small radius ($R_{\mathrm{t}}=9.79a$) with large spontaneous curvatures $C_0$.
$C_0R_{\mathrm{t}}=4$ and $\kappa_1/\kappa_0=4$.
The corresponding movie is provided in Supplementary Movie S3.  }
\end{figure}

First, we investigated the effects of mechanochemical coupling on propagating waves in deformable tubular membranes with a small radius ($R_{\mathrm{t}}= 9.79a$). 
A straight wave propagates at a small spontaneous curvature $C_0$, similar to non-deformable tubes, as the coupling effect is weak (see Fig.~\ref{Fig:tube_snap1}(a) and Supplementary Movie S1). However, undulations of the membrane shape and wavefront are observed at an intermediate $C_0$ (see Fig.~\ref{Fig:tube_snap1}(b) and Supplementary Movie S2). At $C_0R_{\mathrm{t}}=3$ and $\kappa_1/\kappa_0=3$, the tube is deformed in to a meandering form, as shown in the first snapshot of Fig.~\ref{Fig:tube_snap1}(b), owing to the instability of cylindrical shapes. When a wave is stimulated in this meandering tube, the wave propagates on the tube. Owing to the tube deformation, the shape of the protein wave is deformed, which could also affect the membrane shape. 
In addition, the waves are unstable and eventually disappear at large $C_0$ (see Fig.~\ref{Fig:tube_snap2} and Supplementary Movie S3 for $C_0R_{\mathrm{t}}=4$ and $\kappa_1/\kappa_0=4$).
These results indicate that the interaction between membrane deformation and propagating waves changes the stability of the membrane shapes and protein waves.

Figure~\ref{Fig:move1} shows the movement of the center of mass of the waves ($g_{\mathrm{wave},x}$) along the tube ($x$) axis. 
The wave moves almost linearly for small spontaneous-curvature conditions (purple and green lines in Fig.~\ref{Fig:move1}); however, the wave is weakly wobbled by the membrane deformation.
The inputs from the mechanochemical effect on the reaction of $u$ for $H=1/2R_{\mathrm{t}}$ are $-G\partial f_{\mathrm{cv}}/\partial u = 0.0002$ and $-0.0023$ for the purple and green lines in Fig.~\ref{Fig:move1}, respectively.
These results indicate that the waves move at an approximate constant speed even on deformable tubes. The wave speed decreases with decreasing mechanochemical inputs $-G\partial f_{\mathrm{cv}}/\partial u$, as in the case of non-deformable tubes (Fig.~\ref{Fig:3Dwave}).

At large $C_0$ and $\kappa_1/\kappa_0$ values, $g_{\mathrm{wave},x}$ moves slightly but the wave disappears (Fig.~\ref{Fig:tube_snap2}).
However, at $C_0R_{\mathrm{t}}=4$ and $\kappa_1/\kappa_0=4$, the mechanochemical input $-G\partial f_{\mathrm{cv}}/\partial u|_{H=1/2R_{\mathrm{t}}} = -0.0073$, which is still in the range of the stable wave for the non-deformable tubular membranes (see Fig.~\ref{Fig:3Dwave}). Therefore, the disappearance of the wave is partly caused by the reduction in diffusion through the narrow neck.

\begin{figure}[!tb]
	\centering
	\includegraphics[width=8cm]{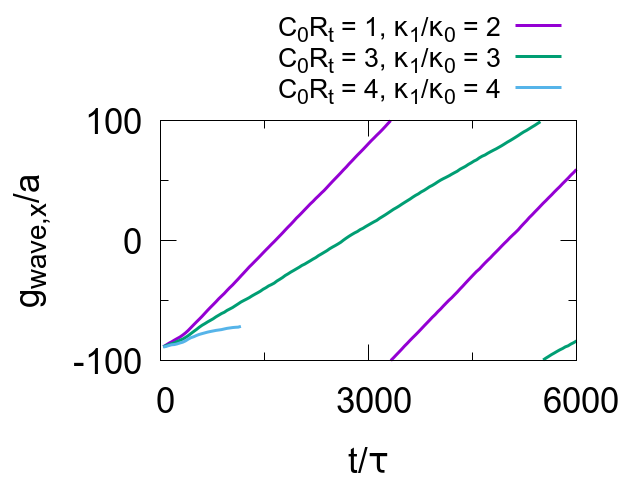}
	\caption{\label{Fig:move1}
		Time evolution of $x$-coordinate of the center of mass of the wave ($g_{\mathrm{wave},x}$), on deformable tubular membranes with the small radius ($R_{\mathrm{t}}=9.79a$).
		The simulation condition is the same as that of Figs.~\ref{Fig:tube_snap1} and \ref{Fig:tube_snap2}.
	}
\end{figure}

\begin{figure}[!tb]
	\centering
	\includegraphics[width=8cm]{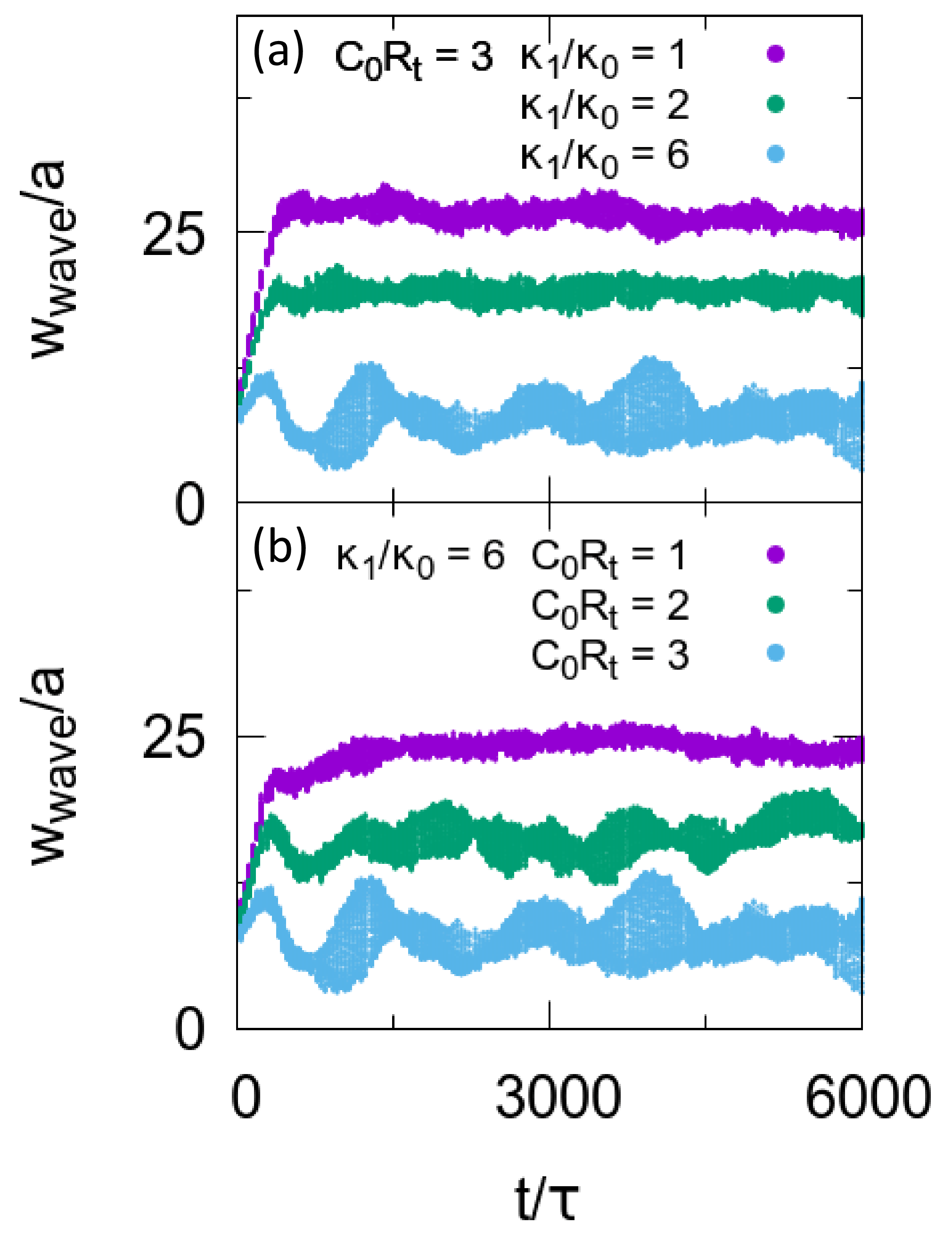}
	\caption{\label{Fig:width}
		Time evolution of the wave width on deformable tubular membranes with a small radius ($R_{\mathrm{t}}=9.79a$).
		(a) $\kappa_1/\kappa_0$ varied at $C_0R_{\mathrm{t}}=3$. (b) $C_0R_{\mathrm{t}}$ varied at $\kappa_1/\kappa_0=6$.
	}
\end{figure}

The wave width also decreases with decreasing mechanochemical input $-G\partial f_{\mathrm{cv}}/\partial u$ (Fig.~\ref{Fig:width}).
Figure~\ref{Fig:width}(a) shows the time evolution of the wave width at $C_0R_{\mathrm{t}}=3$ for $\kappa_1/\kappa_0=1$, $2$, and $6$ ($-G\partial f_{\mathrm{cv}}/\partial u = -0.0006$, $-0.0015$, and $-0.0048$ at $H=1/2R_{\mathrm{t}}$, respectively). Straight waves propagate at $\kappa_1/\kappa_0 = 1$ and $2$. The wave widths are also approximately the same everywhere and maintain a constant value in time.
Similar changes were observed at $\kappa_1/\kappa_0 = 6$ for $C_0R_{\mathrm{t}} = 1$, $2$, and $3$ ($-G\partial f_{\mathrm{cv}}/\partial u = 0.0002$, $-0.001$, and $-0.0048$ at $H=1/2R_{\mathrm{t}}$, respectively) (Fig.~\ref{Fig:width}(b)).

\begin{figure}[!tb]
  \centering
  \includegraphics[width=8cm]{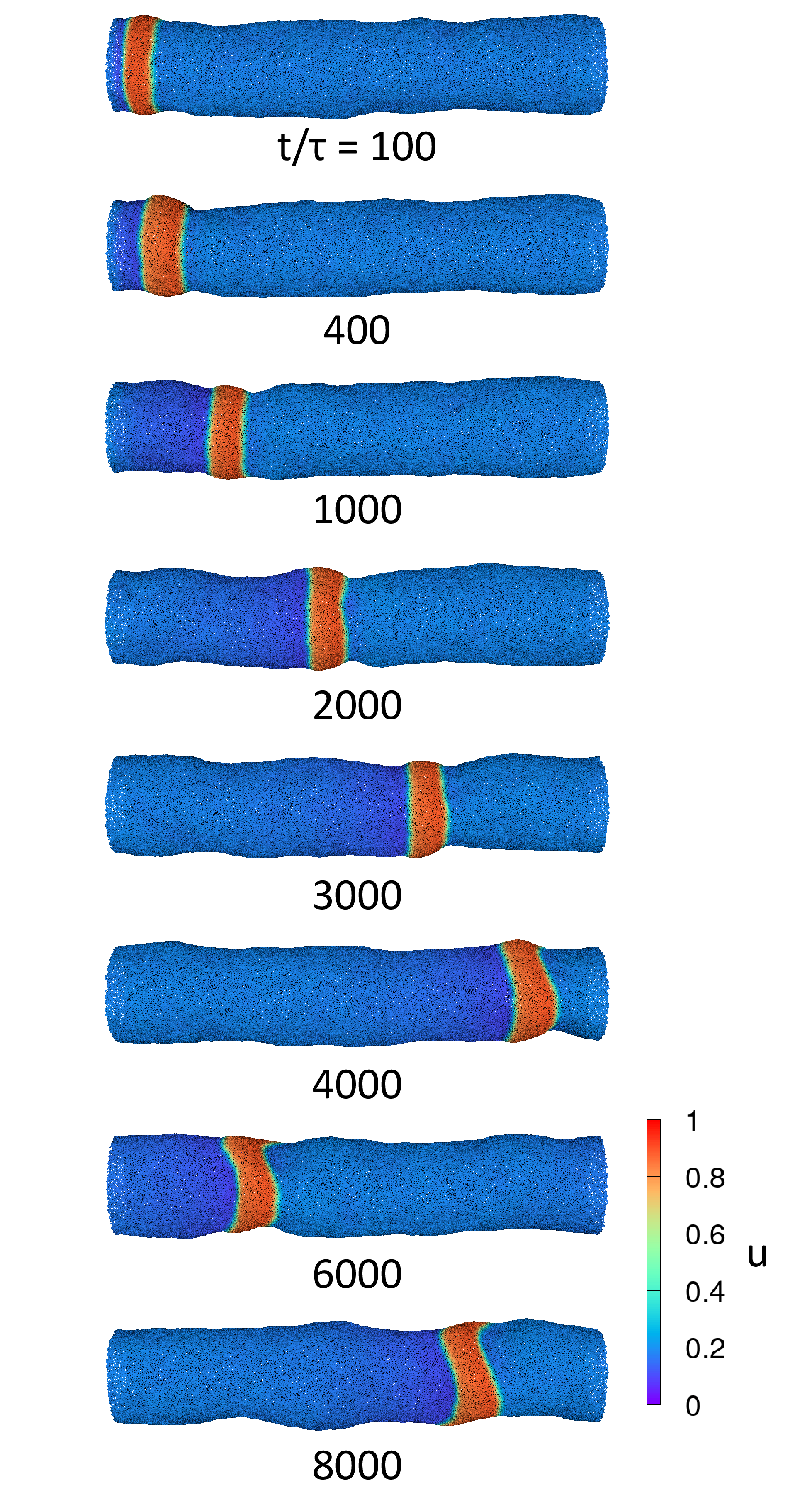}
  \caption{\label{Fig:tube_snap3}
    Typical sequential snapshots of propagating waves on deformable wide tubes ($R_{\mathrm{t}}=19.6a$) at $C_0R_{\mathrm{t}}=4$ and $\kappa_1/\kappa_0=4$. The color variation indicates different concentration of the curvature-inducing protein, $u$. The corresponding movie is provided in Supplementary Movie S4.
  }
\end{figure}

\begin{figure}[!tb]
  \centering
  \includegraphics[width=8cm]{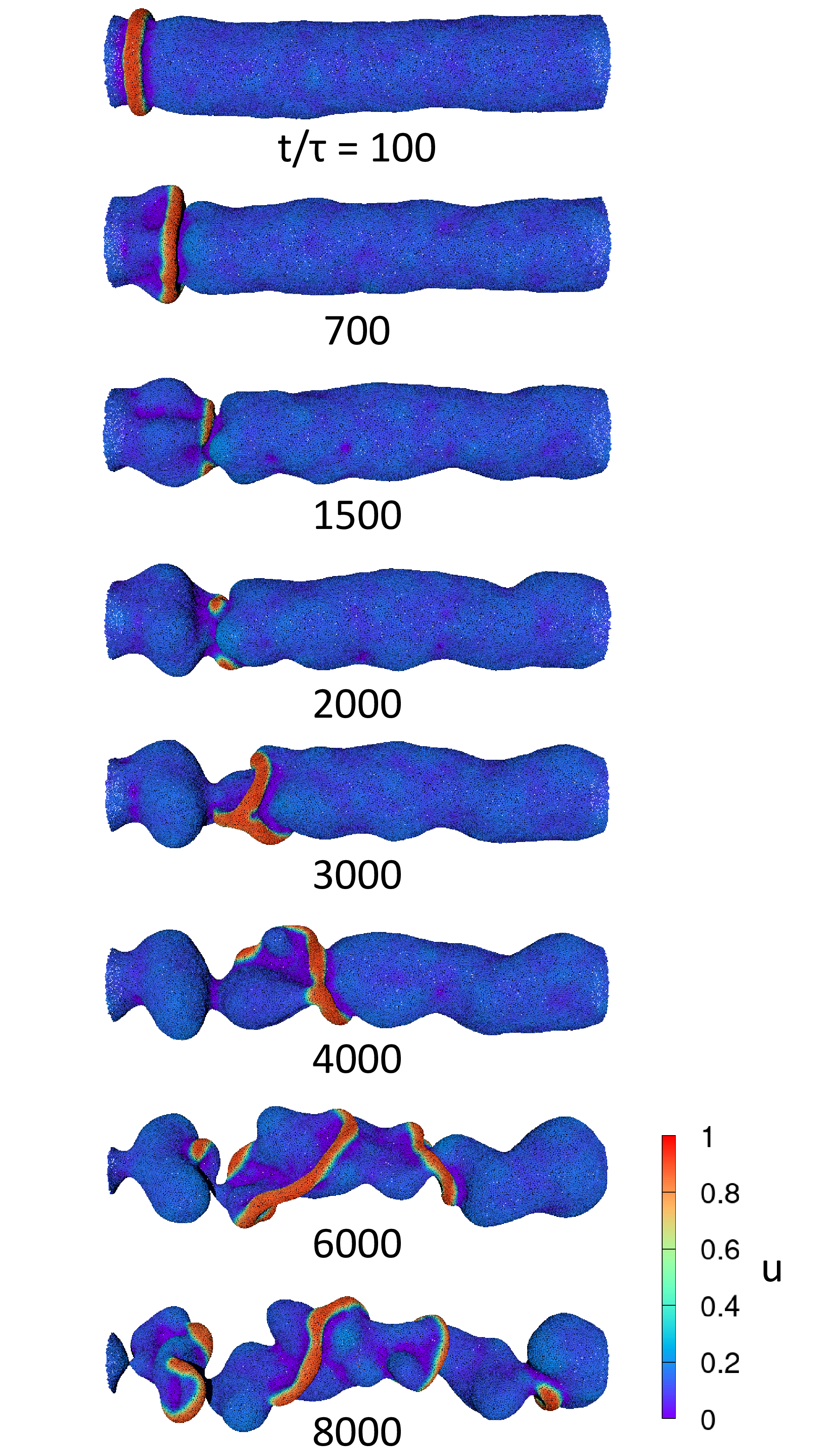}
  \caption{\label{Fig:tube_snap4}
    Typical sequential snapshots of propagating waves on deformable wide tubes ($R_{\mathrm{t}}=19.6a$) at $C_0R_{\mathrm{t}}=8$ and $\kappa_1/\kappa_0=8$. The color variation indicates different concentration of the curvature-inducing protein, $u$. The corresponding movie is provided in Supplementary Movie S6.
  }
\end{figure}

Mechanochemical effects were also observed for the wide tubes (Figs.~\ref{Fig:tube_snap3} and \ref{Fig:tube_snap4}). At $C_0R_{\mathrm{t}} = 4$ and $\kappa_1/\kappa_0=4$, the wavefront is undulated with a small azimuthal deformation of the membrane. In contrast, the other tube regions maintain a straight shape (see Fig.~\ref{Fig:tube_snap3} and Supplementary Movie S4). The wavefront undulation also occurs even for small $C_0$ and large $\kappa_1/\kappa_0$ (see Supplementary Movie S5)). The detection and evaluation methods of the wavefront undulation are described in Appendix~\ref{apppen}.
A large membrane deformation is induced when the value of $C_0$ is large, and the wave locally disappears and breaks due to the winding membrane shape (Fig.~\ref{Fig:tube_snap4}, $t/\tau = 1500\text{--}2000$, see also Supplementary Movie S6). Once the wave breaks, the shapes of the wave and membrane become largely disordered (Fig.~\ref{Fig:tube_snap4}, $t/\tau = 3000\text{--}8000$).

Figure~\ref{Fig:move2} shows the movements of the $x$-coordinate of the vertices in the wave ($r_{\mathrm{wave},x}$) on a deformable wide tube ($R_{\mathrm{t}}=19.6a$) at $C_0R_{\mathrm{t}}=8$ and $\kappa_1/\kappa_0=8$. The wave breaks and splits into two or three pieces (Fig.~\ref{Fig:move2}, $t/\tau = 3500\text{--}7000$).
Even though the shapes of the waves and membranes are disordered, the waves move almost linearly along the membrane tube after division.

\begin{figure}[!tb]
	\centering
	\includegraphics[width=8cm]{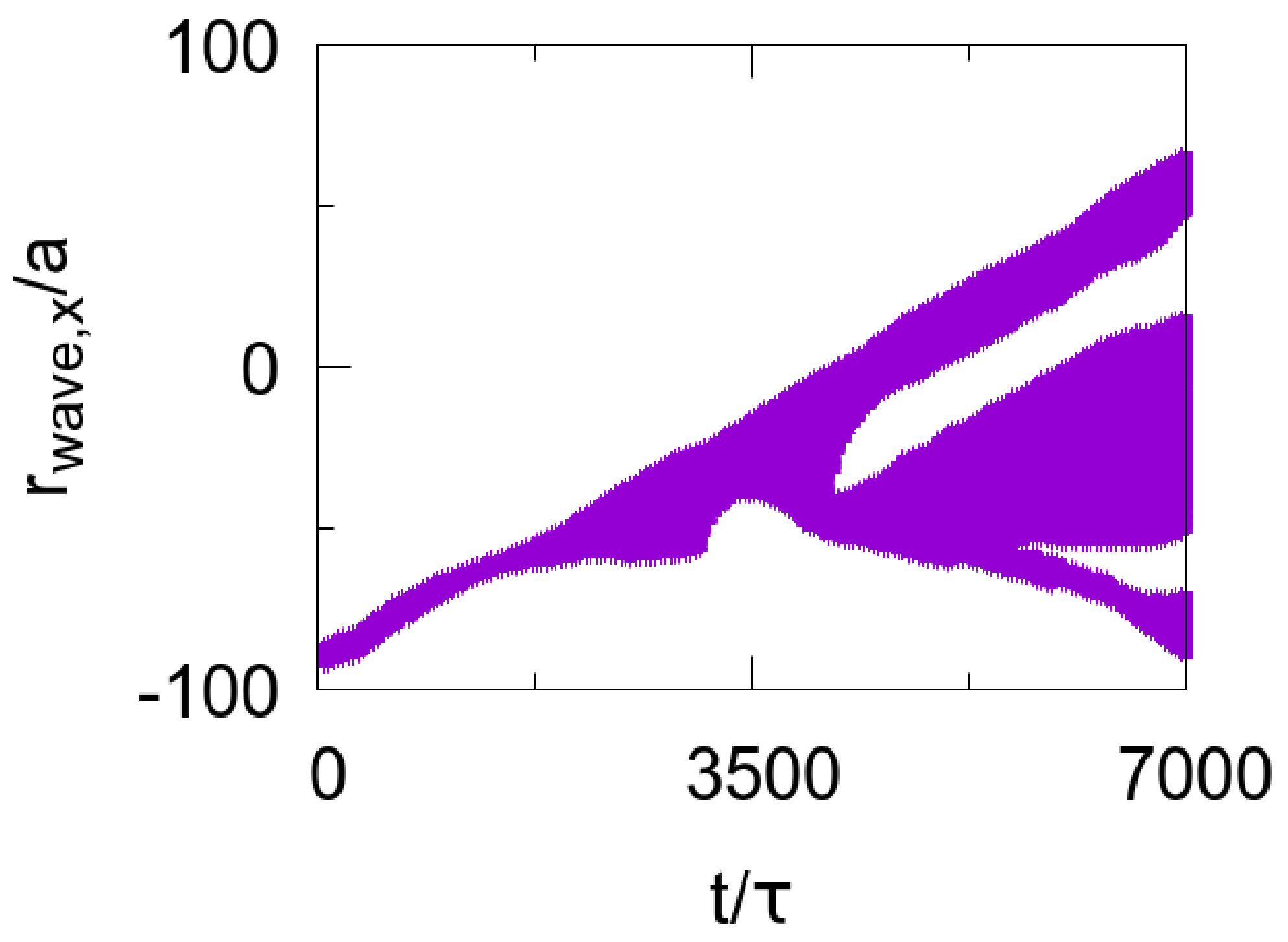}
	\caption{\label{Fig:move2}
		Time evolution of the $x$-coordinate of vertices in the wave ($r_{\mathrm{wave},x}$) on a deformable wide tube ($R_{\mathrm{t}}=19.6a$) at $C_0R_{\mathrm{t}}=8$ and $\kappa_1/\kappa_0=8$.
		The simulation condition is the same as that of Fig.~\ref{Fig:tube_snap4}.
	}
\end{figure}

\begin{figure}[!tb]
  \centering
  \includegraphics[width=8cm]{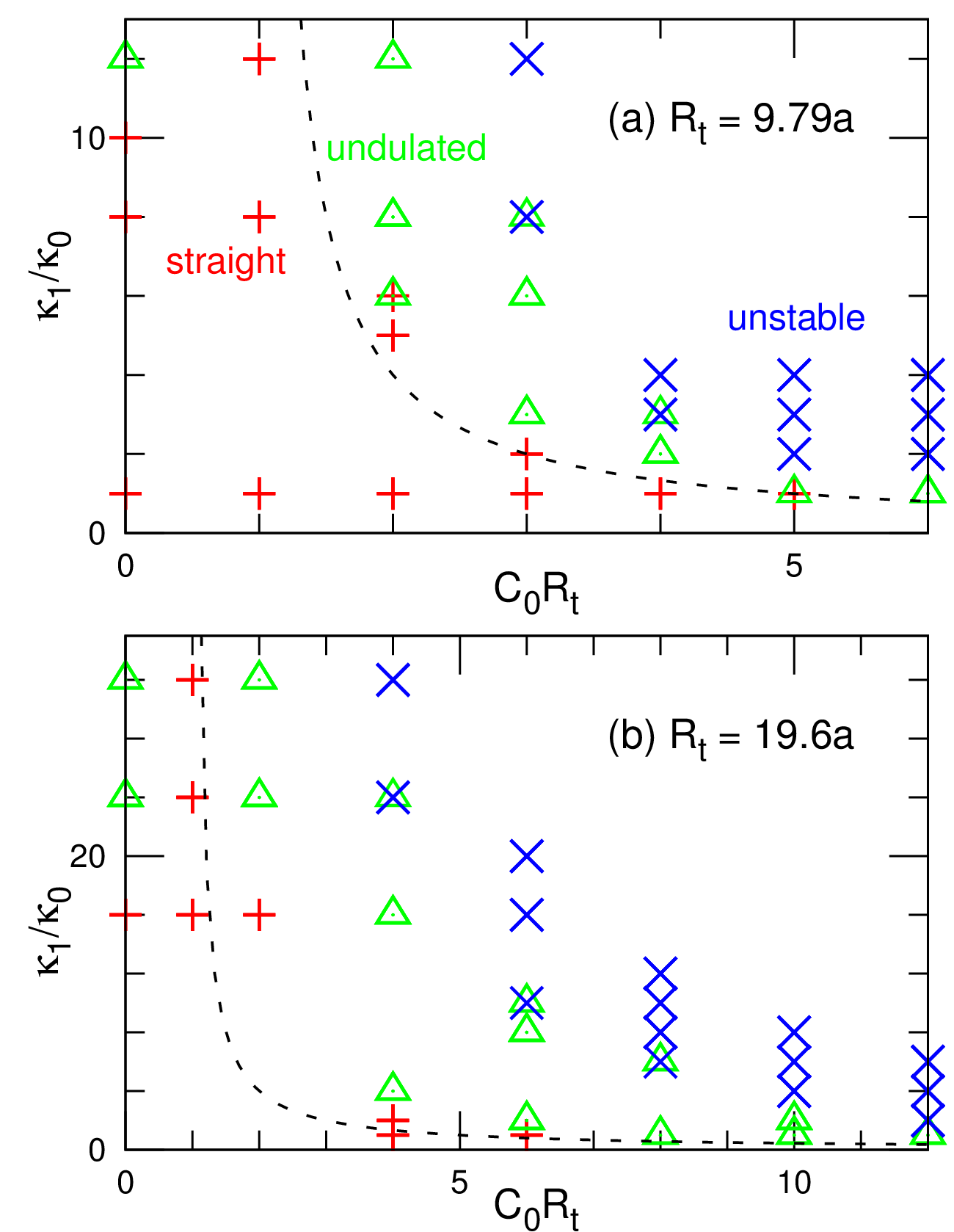}
  \caption{\label{Fig:tube_phase}
    Pattern diagrams with deformable tubular membranes for $R_{\mathrm{t}}=9.79a$ and $19.6a$ (narrow tube and wide tube, respectively). The symbols represent the simulation results. Overlapped symbols indicate the coexistence of multiple patterns.
The dashed lines are given by Eq.~(\ref{eq:fzero}) with $u=0.2$.
  }
\end{figure}

Based on these results, we classified the waves into three phases, straight, undulated, and unstable, as shown in Fig.~\ref{Fig:tube_phase}.
The shapes of the tubes and waves remain straight in the straight wave phase, similar to the case of non-deformable tubes (see Figs.~\ref{Fig:tube_snap0} and \ref{Fig:tube_snap1}(a)). This phase is observed when the effect of the mechanochemical coupling is weak.
In the unstable wave phase at large  $C_0$,
the waves disappear and break. At large $\kappa_1/\kappa_0$, the wave phase appears at smaller values of $C_0$,
since the mechanochemical coupling is strong. Large membrane deformations, such as the formation of narrow necks, are also induced due to large values of $C_0$ (see Fig.~\ref{Fig:tube_snap2}).
The wave's shape is disordered when the wave breaks. When the disordered wave does not disappear, the tube's shape is largely deformed due to the disordered wave (see Fig.~\ref{Fig:tube_snap4}).
In the intermediate condition, tube shape and/or the wavefront is undulated but the wave propagates. We call it undulated wave phase.
The membrane shape and wavefront are undulated and unstable, depending on the spontaneous curvature, bending rigidity, and tube radius.
At $C_0=0$, a cylindrical tube is stable at any value of $\kappa_1/\kappa_0$ in thermal equilibrium.
However, the wave deforms the tube in azimuthal direction at large $\kappa_1/\kappa_0$,  as explained in Sec.~\ref{sec:azi}.

\subsection{Analysis of undulated waves}\label{sec:azi}
\begin{figure}[!tb]
  \centering
  \includegraphics[width=8cm]{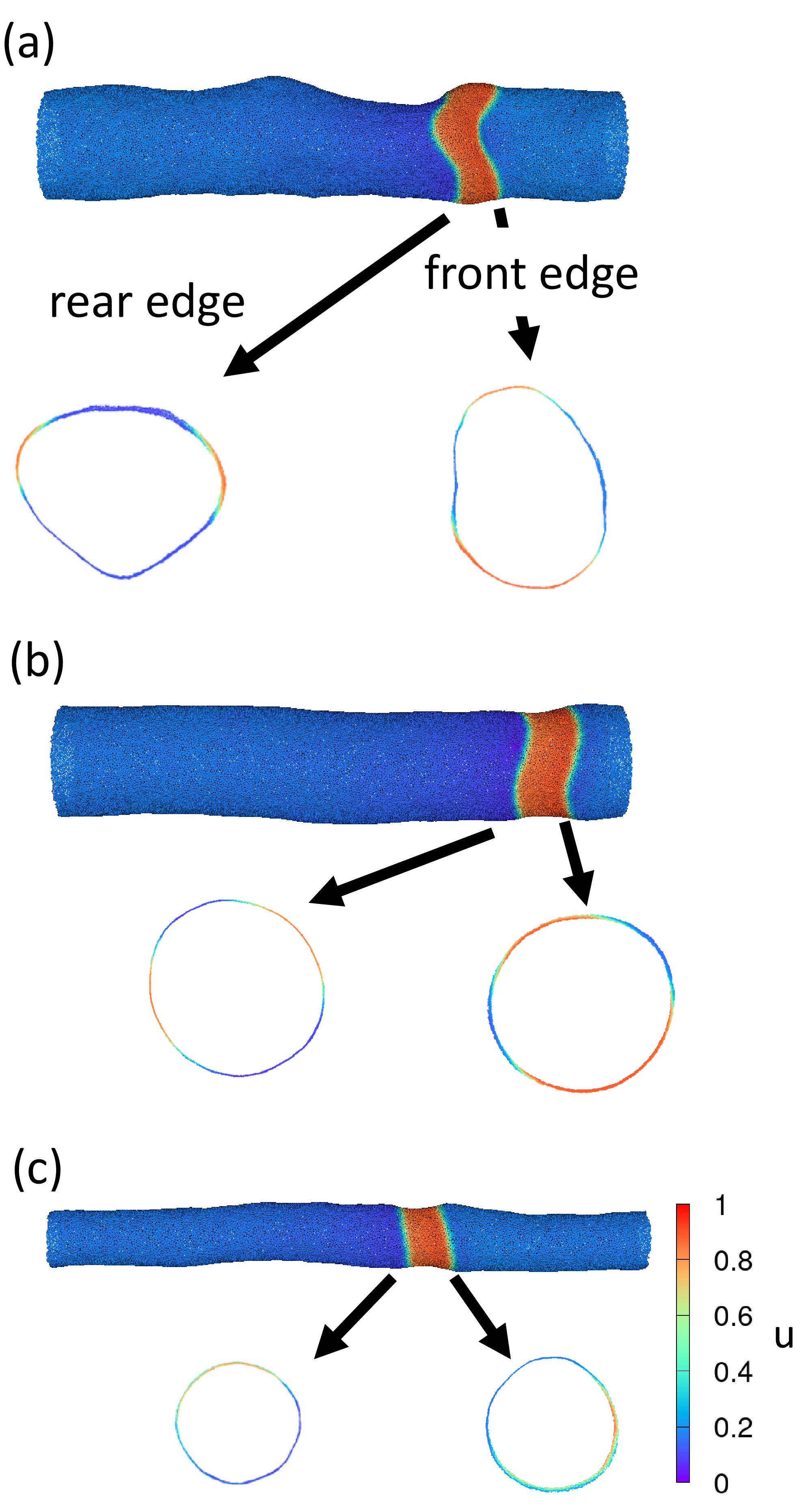}
  \caption{\label{Fig:tube_snap5}
    Typical snapshots of undulated waves and cross-section views of the waves at front and rear edges. (a) $C_0R = 4$, $\kappa_1/\kappa_0=4$, and $R_{\mathrm{t}}= 19.6a$ (wide tube).
(b) $C_0R_{\mathrm{t}} = 0$ and $\kappa_1/\kappa_0=32$, and $R_{\mathrm{t}}= 19.6a$ (wide tube).
(c) $C_0R_{\mathrm{t}} = 0$ and $\kappa_1/\kappa_0=12$, and $R_{\mathrm{t}}= 9.8a$ (narrow tube) 
The color variation indicates different concentration of the curvature-inducing proteins, $u$.
  }
\end{figure}

For further investigation, we focused on the undulated wave phase (green triangles in Fig.~\ref{Fig:tube_phase}). Figure~\ref{Fig:tube_snap5} shows typical snapshots of the undulated waves and their cross-section views. For a wide tube ($R_{\mathrm{t}}= 19.6a$), at $C_0R_{\mathrm{t}} = 4$ and $\kappa_1/\kappa_0 = 4$, the global shape of the tube is not undulated, but the wavefront is winding (Fig.~\ref{Fig:tube_snap5}(a)). The front and rear edges of the wave elongate perpendicularly (major axes of their elliptical shapes at the cross sections are perpendicular). Curvature-inducing proteins exist in two separate regions in the cross-sectional view at the edges of the wave owing to this wavefront shape, which induce azimuthal deformation in the membrane shape. In the azimuthal spatial distribution, the dominant azimuthal wavenumber is two.
At large $\kappa_1/\kappa_0$ and $C_0=0$, a winding wavefront appears (e.g. at $C_0R_{\mathrm{t}} = 0$ and $\kappa_1/\kappa_0 = 32$, see Fig.~\ref{Fig:tube_snap5}(b) and Supplementary Movie S5).
In contrast, for a narrow tube ($R_{\mathrm{t}}= 9.79a$), such winding wavefront with the locally elliptic tube shape are not observed. At $C_0R_{\mathrm{t}} = 0$ and $\kappa_1/\kappa_0 = 12$, the wavefront is winding (Fig.~\ref{Fig:tube_snap5}(c)). In this condition, the dominant azimuthal wavenumber of protein distribution is one.
			
We focused on the wavefront of the propagating wave and performed a discrete Fourier transformation (DFT) on the cross-sections at the wavefront to quantify the azimuthal deformation ($R_{\mathrm{t}}= 19.6a$, $C_0R_{\mathrm{t}} = 4$, and $\kappa_1/\kappa_0 = 4$).
$A_{u_q}$ and $A_{H_q}$ indicate the $q$th amplitudes of DFT for the concentration of the curvature-inducing protein $u$ and smoothed local membrane curvature $\tilde{H}$ in an azimuthal direction, respectively.
The magnitude of undulation of the wavefront is quantified by the value $\langle d_{\mathrm{RMSD}} \rangle$, which is the average of root mean squared deviations between the wavefront and local geodesic lines, described in Appendix~\ref{apppen}. Figure~\ref{Fig:tube_DFT} shows that the undulation of the wavefront increases, and the $q=2$ modes of $u$ and $\tilde{H}$ dominantly grow together.

\begin{figure}[!tb]
  \centering
  \includegraphics[width=8cm]{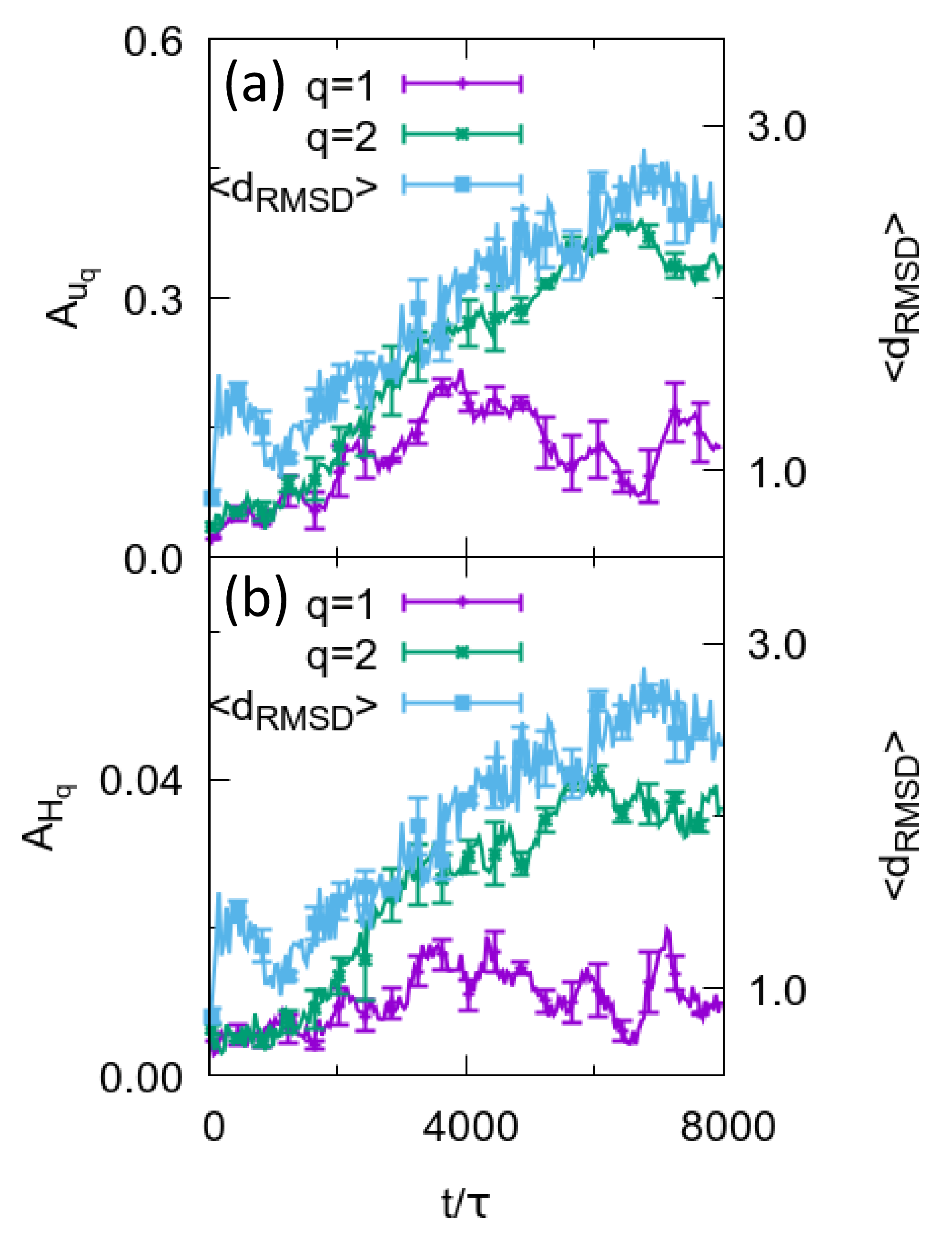}
  \caption{\label{Fig:tube_DFT}
    Time evolution of the magnitude of undulation of wavefronts, $\langle d_{\mathrm{RMSD}} \rangle$ for $C_0R_{\mathrm{t}} = 4$, $\kappa_1/\kappa_0=4$, and $R_{\mathrm{t}}= 19.6a$ (wide tube), with $q$th amplitudes of DFT for (a) $u$ and (b) $\tilde{H}$ in an azimuthal direction, $A_{u_q}$ and $A_{H_q}$, respectively ($q=1$ and $2$). The simulation condition is the same as that of Figs.~\ref{Fig:tube_snap3} and \ref{Fig:tube_snap5}(a).
  }
\end{figure}

It is considered that the azimuthal wavenumber is determined by a combination of the spatial structures of the reaction-diffusion waves and tubular membrane deformation through mechanochemical coupling. 
For the wide tube, the circumference is long enough to form the winding wavefront with azimuthal wavenumber $q=2$. The tubes deform in to the locally elliptic shape.
For the narrow tube, the winding wavefront with azimuthal wavenumber $q=1$ is formed.

\section{Conclusions}
In this study, we have studied the coupled dynamics between the deformation of tubular membranes and excitable reaction-diffusion waves of curvature-inducing proteins. 
Three dynamic modes appear depending on spontaneous curvature, bending rigidity, and tube radius: the straight wave phase with small membrane deformation, undulated wave phase, and unstable wave phase.
Membrane deformation changes the protein-binding rate and diffusion along the tube axis, and they modify the wave shape.
The waves disappear at narrow membrane necks in the unstable phase.

The waves with azimuthal wavenumber $q=2$ (twofold rotationally symmetric) grow with the undulation of the wavefront for the wide tubes, in which the membrane locally deforms in an elliptic shape, and the front and rear edges of the wave elongate perpendicularly. 
Three (or more) -fold symmetric waves may propagate for much wider tubes.
Although axisymmetric assumption has often been used for membrane simulations, 
full three-dimensional simulations are needed to investigate non-axisymmetric waves, as demonstrated in this study.

The shape of organelles is dynamically regulated in living cells.
For example, the stack structure of the Golgi apparatus is formed by the cis side and deconstructed from the opposite side~\cite{Matsuura1,Papanikou1}. In this dynamic equilibrium state, curved membrane structures such as the cisternae and tubular networks are regulated by various proteins at different maturation stages \cite{Kulkarni-Gosavi1}.
Pearled tubular membranes are formed as transport intermediates for cargo delivery from ER to Golgi. The formation of these structures is associated with COPI and COPII \cite{Weigel1,Zucker1}. In addition, the COPI travels during the cargo delivery \cite{Weigel1}.
Reaction-diffusion waves may play a role in these structural changes.
Further, endocytosis in hotspots (emerging from the same position) is enhanced in the presence of propagating waves of curvature-inducing proteins \cite{YangY1}. The propagation of membrane undulations of a certain wavenumber with protein propagation waves might be related to the mechanism of these spatial pattern generation.
We employed the FitzHugh--Nagumo model to generate an excitable chemical wave.
However, our simulation method can be applied to other chemical reaction models
and used to investigate various biomembrane phenomena.

This study considered the proteins that induce laterally isotropic spontaneous curvatures.
BAR superfamily proteins and other proteins have laterally anisotropic shapes (often modeled as an elliptic shape) and bend the bound membrane anisotropically~\cite{Baumgart2,Suetsugu1}; they
exhibit different behaviors from isotropic proteins.
The BAR proteins induce membrane tubulation~\cite{Baumgart2,Suetsugu1,Mim1,Kabaso1,Noguchi5}, 
whereas the isotropic proteins induce the formation of spherical buds~\cite{Hurley1,Foret1,Goutaland1,Noguchi6}.
The maximum binding (sensing) curvature of anisotropic proteins depends on the binding chemical potential, whereas that of isotropic proteins is constant~\cite{Noguchi8,Noguchi9}.
Chemical waves involving BAR proteins have been observed~\cite{WuZ1,YangY1,XiaoS1},
but the effects of anisotropy in the waves are not understood so far.
It is essential to further studies how such differences in curvature-inducing proteins change coupling with reaction-diffusion dynamics.

\begin{figure}[!tb]
  \centering
  \includegraphics[width=8cm]{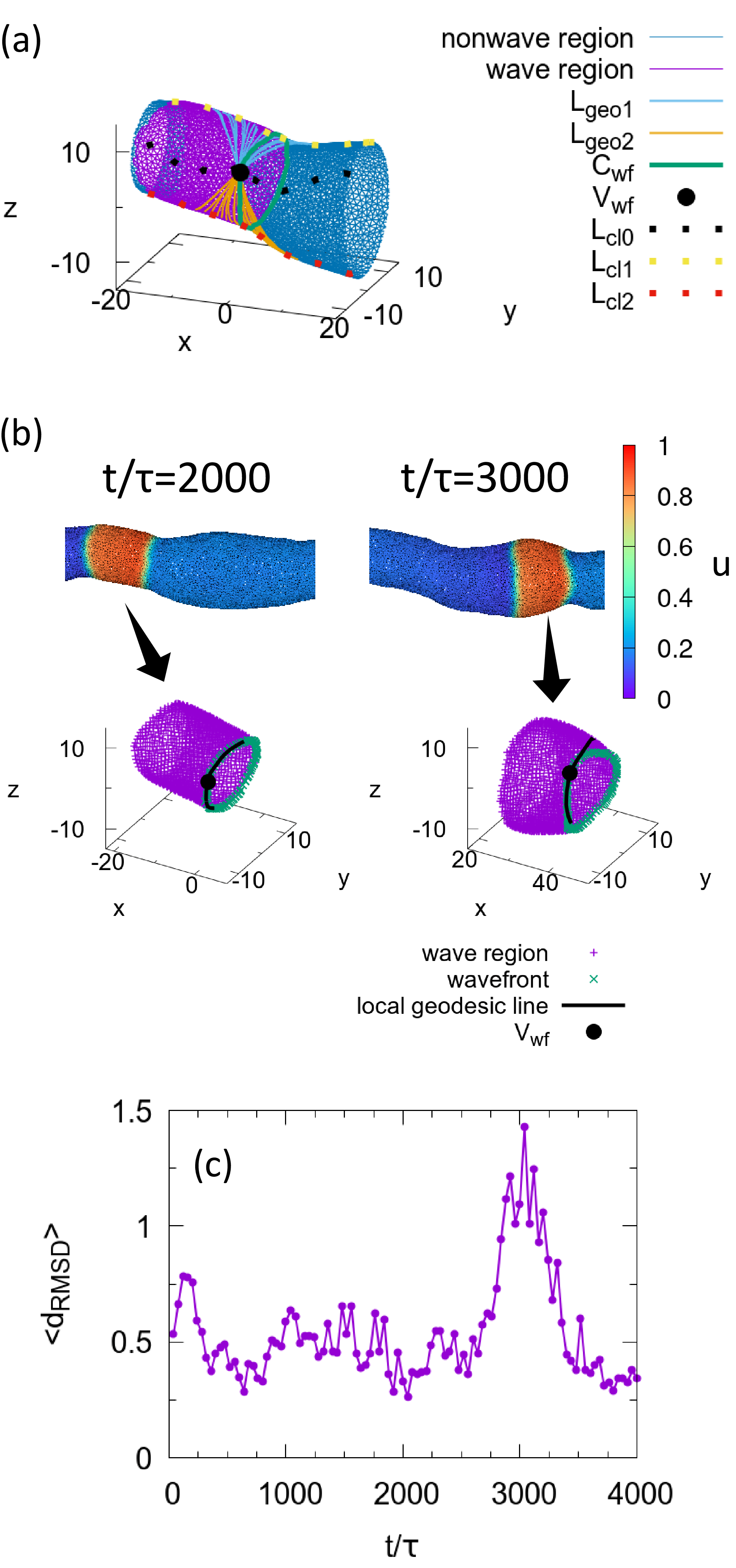}
  \caption{\label{Fig:geodesics}
    Calculation of local geodesic lines and $\langle d_{\mathrm{RMSD}} \rangle$ for a narrow tube ($R_{\mathrm{t}}=9.79a$, $C_0R_{\mathrm{t}}=2$, and $\kappa_1/\kappa_0 = 6$). (a) Lines and points obtained in the calculation process of the local geodesic line. (b) Snapshots of waves and local geodesic lines at $t/\tau = 2000$ and $t/\tau = 3000$. The color in snapshots indicates the concentration of curvature-inducing proteins, $u$. (c) Time development of $\langle d_{\mathrm{RMSD}} \rangle$.}
\end{figure}
	
\begin{acknowledgments}
We would like to thank S. Ishihara and M. Yanagisawa for valuable discussions. 
This work was supported by JSPS KAKENHI Grant Number JP21K03481.
\end{acknowledgments}

\begin{appendix}

\section{Algorithm to evaluate the wavefront undulation}\label{apppen}
First, we determine that a spatial wave exists if $u_{\mathrm{max}} \geq 0.55$ and $u_{\mathrm{max}} - u_{\mathrm{min}} > 0.4$, where $u_{\mathrm{max}}$ and $u_{\mathrm{min}}$ are the maximum and minimum values of $u$, respectively. Vertices for $u \geq 0.55$ are classified to the wave region when the wave exists, and the other vertices to the non-wave region. When vertices in the wave and non-wave regions are adjacent to each other, the internally dividing point, in which $u=0.55$ on the bond between the vertices, is considered a wave edge.
The wavefront is determined by the condition on $v$ at the edge, $v < v_{\mathrm{min}} + 0.4(v_{\mathrm{max}} - v_{\mathrm{min}})$, where $v_{\mathrm{max}}$ and $v_{\mathrm{min}}$ are the maximum and minimum values of $v$, respectively, because when the reaction-diffusion wave propagates, the oscillation phase of $v$ should shift from that of $u$ toward the reverse direction of wave propagation.
		
Let $V_{\mathrm{wf}}$ be a vertex on the wavefront (black point in Fig.~\ref{Fig:geodesics}(a)). Since $V_{\mathrm{wf}}$ is on a bond of the original triangular network graph, it faces two triangles and is adjacent to the two other vertices of the wave edge. Let $L_{\mathrm{wf}}$ be a line connecting those two vertices. We roughly estimate the direction of the line tangent to the wavefront at $V_{\mathrm{wf}}$ as the direction of $L_{\mathrm{wf}}$.
Next, we calculate the cutting line of the tube along the $x$-axis through $V_{\mathrm{wf}}$ using a greedy algorithm to advance in the $x$-axis direction with a lookahead search with a depth of ten on a discrete graph (black dotted line ($L_{\mathrm{cl0}}$) in Fig.~\ref{Fig:geodesics}(a)).
Then, we determine the shortest path from $V_{\mathrm{wf}}$ to $V_{\mathrm{wf}}'$ (the corresponding vertex to $V_{\mathrm{wf}}$ duplicated by cutting) on the graph cut by $L_{\mathrm{cl0}}$. Thus, we obtain a path that starts from $V_{\mathrm{wf}}$, circles the tube, and returns to itself (green circle line ($C_{\mathrm{wf}}$) in Fig.~\ref{Fig:geodesics}(a)).
As well as $L_{\mathrm{cl0}}$, we obtain two cutting lines $L_{\mathrm{cl1}}$ and $L_{\mathrm{cl2}}$, which are through vertices where $C_{\mathrm{wf}}$ is divided by one quarter and three quarters, respectively (yellow and red dotted lines in Fig.~\ref{Fig:geodesics}(a)).
Then, we select several vertices on $L_{\mathrm{cl1}}$ and $L_{\mathrm{cl2}}$ ($V_{\mathrm{cl1}}$ and $V_{\mathrm{cl2}}$) and calculate the approximate shortest path on the graph from $V_{\mathrm{cl1}}$ and $V_{\mathrm{cl2}}$ to $V_{\mathrm{wf}}$ ($L_{\mathrm{geo1}}$ and $L_{\mathrm{geo2}}$, depicted by cyan and orange lines in Fig.~\ref{Fig:geodesics}(a)). The algorithm for obtaining geodesic lines is an iterative refinement of the discrete graph and Dijkstra's algorithm, the details of which can be found in Ref.~\cite{Kanai1}.
We find the best two geodesic lines, one from $L_{\mathrm{geo1}}$ and one from $L_{\mathrm{geo2}}$, which are (i) smoothly connected at $V_{\mathrm{wf}}$, and (ii) nearly parallel to $L_{\mathrm{wf}}$. These local geodesic lines are calculated for approximately $20$ points on the wavefront, and the root mean squared deviation between the geodesic line and the wavefront ($d_{\mathrm{RMSD}}$) is calculated as below:
\begin{equation}
  d_{\mathrm{RMSD}} = \sqrt{\frac{1}{N_{\mathrm{g}}} \sum_i \left| \bm{r}_{\mathrm{g},i} - \bm{r}_{\mathrm{wf,c}} \right|^2 },
\end{equation}
where $N_{\mathrm{g}}$ is the number of vertices of the local geodesic line; $\bm{r}_{\mathrm{g},i}$ is a position vector of the $i$th vertex of the local geodesic line; and $\bm{r}_{\mathrm{wf,c}}$ is the position vector of the vertex of the wavefront closest to the $i$th vertex of the local geodesic line.
The average of $d_{\mathrm{RMSD}}$ ($\langle d_{\mathrm{RMSD}} \rangle$) is used to evaluate the wavefront undulation at each time step.
		
Examples of the calculation results are shown in Fig.~\ref{Fig:geodesics}(b) and (c). At $t/\tau = 2000$, the wavefront is not undulated and the local geodesic line is aligned along the wavefront, whereas at $t/\tau = 3000$, the wavefront is undulated and a local geodesic line deviates slightly from the wavefront. (see Fig.~\ref{Fig:geodesics}(b)). Thus, $\langle d_{\mathrm{RMSD}} \rangle$ has a small value at $t/\tau \leq 2000$ but increases around $t/\tau = 3000$ (see Fig.~\ref{Fig:geodesics}(c)). These results demonstrate that the undulation of the wavefront can be evaluated using $\langle d_{\mathrm{RMSD}} \rangle$.

\end{appendix}		


%

\end{document}